\documentclass[12pt]{iopart}
\usepackage{graphicx}

\usepackage{amsfonts}
\usepackage{amssymb} 
\usepackage{mathrsfs}  

\usepackage{url}

\begin{document}

\title{Physics-informed Meta-instrument for eXperiments (PiMiX) with applications to fusion energy}

\author[LANL]{Zhehui Wang\textsuperscript{1}, Shanny Lin\textsuperscript{1,2}, Miles Teng-Levy\textsuperscript{1}, \\
 P. Chu\textsuperscript{1},  B. T. Wolfe\textsuperscript{1}, C. S. Wong\textsuperscript{1}, C. S. Campbell\textsuperscript{1},\\
Xin Yue\textsuperscript{3}, Liyuan Zhang\textsuperscript{4},
D. Aberle\textsuperscript{1}, M. Alvarado Alvarez\textsuperscript{1}, D. Broughton\textsuperscript{1},
Ray T. Chen\textsuperscript{2}, Baolian Cheng\textsuperscript{1}, F. Chu\textsuperscript{1}, \\ Eric R. Fossum\textsuperscript{3},  M. A. Foster\textsuperscript{5}, C.-K. Huang\textsuperscript{1}, 
V. Kilic\textsuperscript{5}, \\ Karl Krushelnick\textsuperscript{6},  W. Li\textsuperscript{1}, Eric Loomis\textsuperscript{1}, T. Schmidt Jr\textsuperscript{1},
Sky K. Sjue\textsuperscript{1}, C. Tomkins\textsuperscript{1},  
 D. A. Yarotski\textsuperscript{1}  \& Renyuan Zhu\textsuperscript{4}}


\address{\textsuperscript{1} Los Alamos National Laboratory, Los Alamos, NM 87545, USA}
\address{\textsuperscript{2} The University of Texas, Austin, Austin, TX 78712, USA}
\address{\textsuperscript{3} Dartmouth College, Hanover, NH 03755, USA}
\address{\textsuperscript{4} California Institute of Technology, Pasadena, CA 91125, USA}
\address{\textsuperscript{5} Johns Hopkins University, Baltimore, MD 21218, USA}
\address{\textsuperscript{6} University of Michigan, Ann Arbor, MI 48109, USA}

\ead{zwang@lanl.gov}
\vspace{10pt}
\begin{indented}
\item[] (Manuscript based on a recent presentation in the 29th IAEA Fusion Energy Conference (FEC), London, UK, Oct. 16 - 21, 2023) 
\end{indented}

\begin{abstract}
Data-driven methods (DDMs), such as deep neural networks, offer a generic approach to integrated data analysis (IDA), integrated diagnostic-to-control (IDC) workflows through data fusion (DF), which includes multi-instrument data fusion (MIDF), multi-experiment data fusion (MXDF), and simulation-experiment data fusion (SXDF). These features make DDMs attractive to nuclear fusion energy and power plant applications, leveraging accelerated workflows through machine learning and artificial intelligence. Here we describe Physics-informed Meta-instrument for eXperiments (PiMiX) that integrates X-ray (including high-energy photons such as $\gamma$-rays from nuclear fusion), neutron and others (such as proton radiography) measurements for nuclear fusion. PiMiX solves multi-domain high-dimensional optimization problems and integrates multi-modal measurements with multiphysics modeling through neural networks. Super-resolution for neutron detection and energy resolved X-ray detection have been demonstrated. Multi-modal measurements through MIDF can extract more information than individual or uni-modal measurements alone.  Further optimization schemes through DF are possible towards empirical fusion scaling laws discovery and new fusion reactor designs.
\end{abstract}

%
%
%
\maketitle
%
%
\tableofcontents

\section{Introduction}
\begin{figure}[htbp] 
  \centering
   \includegraphics[width=5.0in, angle=0]{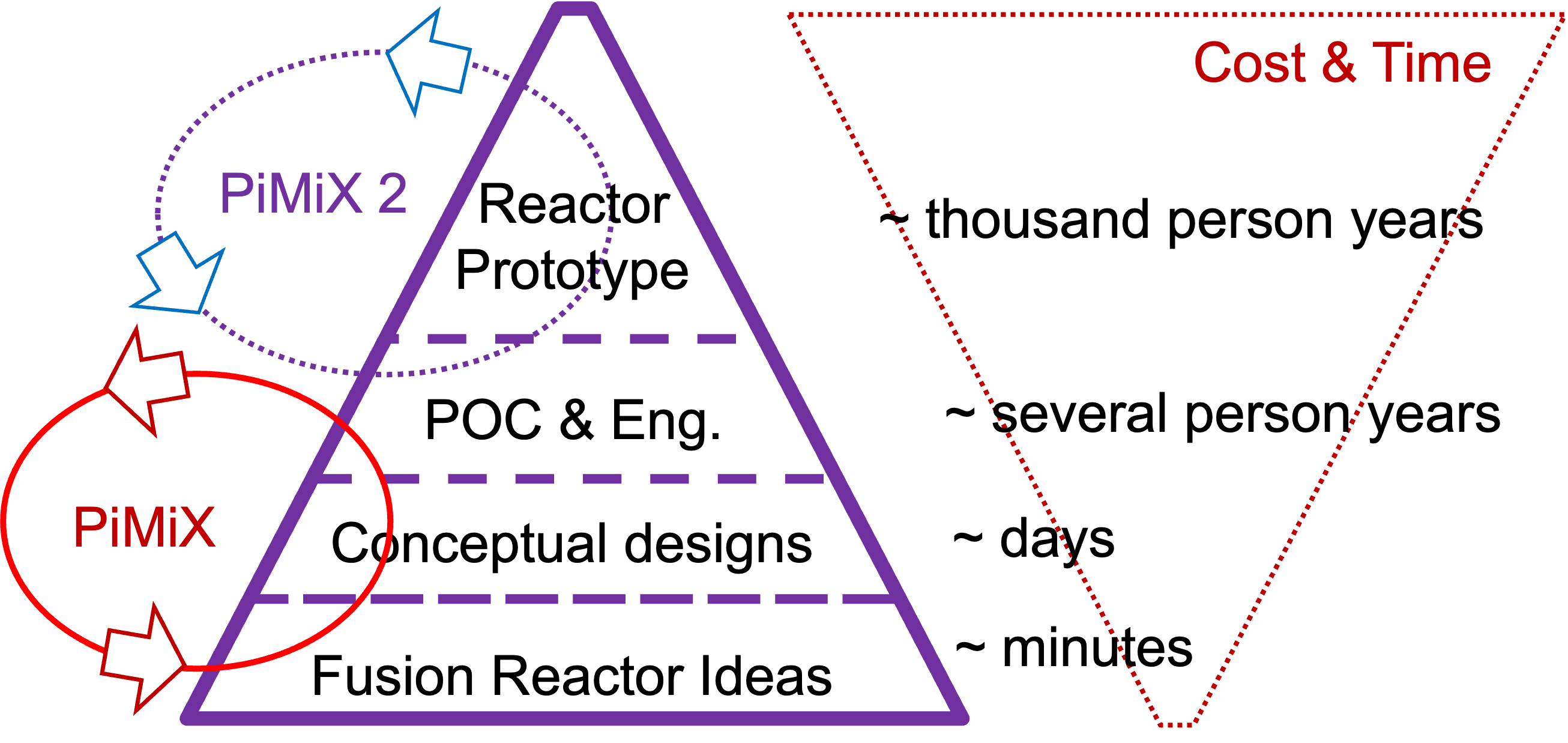} 
   \caption{Nuclear fusion reactor development process can be described by a pyramid starting from many possible ideas to a small number of eventual reactor designs. From the cost and time perspectives, fusion reactor maturation may be regarded as an inverted pyramid shown on the right. PiMiX as a generic heterogeneous data fusion (DF) platform, can be used both for conceptual maturation, as well as optimization of experiments.}
  \label{fig:px1}
 \end{figure}

	Recent controlled release of nuclear fusion energy by laboratory plasmas, such as in the National Ignition Facility (NIF)~\cite{NIF} and in the Joint European Torus (JET)~\cite{JET}, ushered in a new era of nuclear fusion. NIF used a form of inertial confinement fusion (ICF) to generate fusion energy exceeding 3 MJ per shot,  or a fusion-energy-over-laser-energy gain (Q) satisfying $1 <$ Q $< 2$.  JET used a form of magnetic confinement fusion (MCF) to achieve a fusion yield approaching 60 MJ per discharge, or a fusion-power-over-heating-power in steady state reaching 0.5 $<$ Q $<$ 1.  NIF released most of the fusion energy within 1 ns. JET maintained an 10 MW of fusion power, which doubled the previous record, for about 5 seconds. Both NIF and JET plasmas were tritium rich, consuming about 0.1 mg of tritium for 0.07 mg of deuterium in JET, taking advantage of the large deuterium-tritium (DT) fusion cross section. In addition to multi-institutional collaborations such as NIF, JET and ITER, which are supported by public funds, new fusion concepts and prototypes supported by private investments are thriving~\cite{Sur:2022}. There is still plenty of room for innovative fusion energy concepts and improvements over the existing concepts before fusion energy becomes a reliable renewable carbon-neutral power source, with $Q > 10$. Meanwhile, many low-cost thermonuclear fusion concepts today reside in the parameter space corresponding to the fusion triple product ($n_iT\tau_E$) below 10$^{16}$ keV·s/m$^3$, or $Q < 0.1$. Improvement over existing or discovery of new promising concepts with $n_iT\tau_E$ above 10$^{18}$ keV·s/m$^3$ will drastically improve reward/risk ratios for further investment, as illustrated in FIG.~\ref{fig:px1}.
	
 Here we describe a new artificial intelligence (AI) and machine learning (ML)-enhanced measurement and DF  framework, called Physics-informed Meta-instrument for eXperiments (PiMiX) for data collection, data handling and automated data interpretation, in the era of laboratory ignited plasmas. PiMiX, based on neural networks and deep learning algorithms, integrates physics understanding with heterogenous data streams from experiments, multiscale multi-physics simulations, and metadata such as material properties for data interpretation. PiMiX also offers predictive potential for experiments and fusion plasma concepts. Therefore, PiMiX can be used, through the lower feedback loop as illustrated in FIG.~\ref{fig:px1}, for experimental design, optimization and reduction of the turn-around time and cost for new concepts and iterations over a large design space (reactor ideas, for example).
 
 PiMiX uses machine learning algorithms to integrate measurement hardware (X-ray, neutron, proton imaging devices, {\it etc.}), data processing, and data interpretation. Applications of machine learning to plasmas are a relatively new yet rapidly growing interdisciplinary field~\cite{r1, r2}. PiMiX is built upon recent progress in imaging, machine learning from images, and automated processing of images from a diverse setting of plasma experiments~\cite{r3,r4,r5,r6,r7}. Successful applications of deep learning for data analysis such as particle tracking, classification, contour extraction and nonlinear  three-dimensional (3D) reconstruction and regression have been demonstrated. 

 Besides automated and integrated data processing from multiple diagnostics suites, PiMiX intends to address whether, through multiple diagnostic data integration or experimental DF, and other forms of DF such as simulation and experiment DF (SXDF), more information can be extracted than by using individual diagnostics alone. There are ample existing examples of experimental DF, which make use of data complementarity from different measurements. By using X-ray and neutron measurements simultaneously, material identification is possible. By using several imaging cameras from different line of sights, 3D reconstructions of a plasma density distribution have been obtained~\cite{Vol:2017}. By using dual-energy X-ray sources, improvements in imaging contrast and radiation dose have been demonstrated in X-ray computed tomography. PiMiX also leverages the fact that a large number of diagnostics are now or will be available from NIF, JET and ITER for experimental DF. There are more than 100 implosion target diagnostics in NIF~\cite{Kil:2023}, including 37 nuclear, 13 optical and 57 X-ray diagnostics. The diagnostics measure everything between the edge of the implosion down to nuclear interactions. Terabytes of data are generated per shot from NIF’s diagnostic suite. In JET, diagnostic enhancement for DT experiments aimed at better spatial, temporal, and energy
resolution while increasing measurement coverage of neutrons, $\gamma$-rays, fast ions, instabilities signals~\cite{JET:1}. In ITER, a large array of instruments, magnetic, neutron, optical, bolometric, spectroscopic, microwave, infrared, wall and operational diagnostics, will provide the data necessary to control, evaluate and optimize plasma performance in ITER and to further the understanding of plasma physics in the $Q>1$ regime.

Below, we first describe PiMiX DF framework in Sec.~\ref{sec:fw}, and explain the complementarity of traditional physics-driven models (PDMs) and the new DDMs using machine learning algorithms. In Sec.~\ref{sec:hd}, we describe image sensors and scintillator based measurement hardware, motivated by validation of thermonuclear plasma ignition models. The emphasis is on X-ray photon and neutron counting capabilities towards higher information yield from fusion experiments. In Sec.~\ref{sec:dataM} on data and machine learning in PiMiX, we highlight various experimental data sets, synthetic data  sets, deep neural network algorithms, together with a Bayesian inference approach to PiMiX. In Sec.~\ref{ex:val}, we highlight the initial PiMiX results in super spatial resolution for neutron detection, energy-resolved X-ray measurement through physics-informed machine learning, and integration of two diagnostics (proton and X-ray imaging) for higher information yield. Further PiMiX applications for fusion conceptual maturation, such as fusion yield scaling, and experimental optimization, such as higher fusion yield, are possible. 

\section{PiMiX data-fusion framework \label{sec:fw}}

\begin{figure}[htbp] 
  \centering
   \includegraphics[width=5.0in, angle=0]{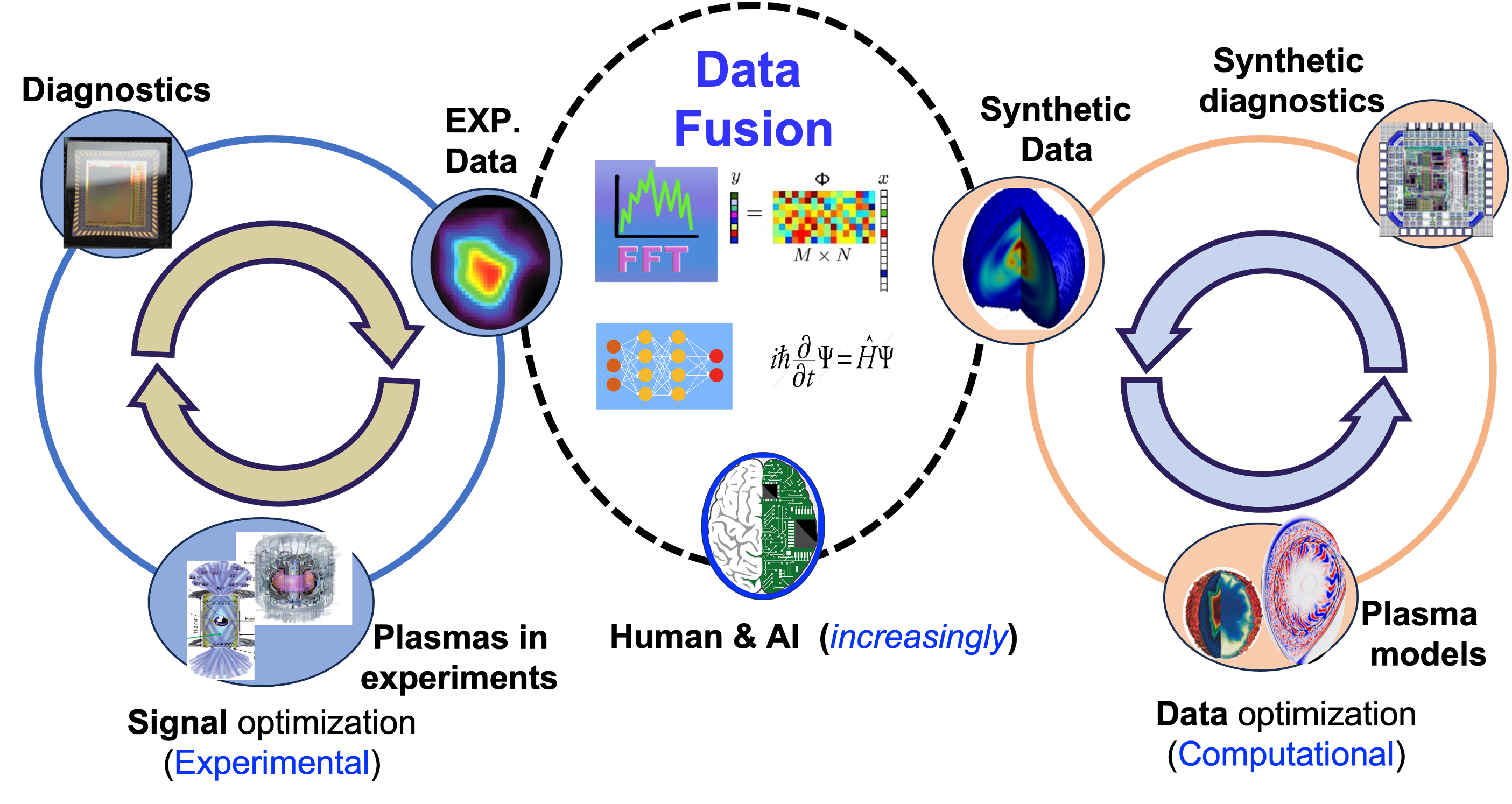} 
   \caption{PiMiX framework integrates data collection (the left loop) with physics and modeling of data (the right loop) through data fusion (the middle loop). The data fusion loop can feed back to the experimental data collection through experimental optimization and real-time controls, as shown in FIG.~\ref{fig:px1}. The data fusion loop can also feed back to the modeling and interpretation to fine tune models. A growing number of analysis tools are now available for data fusion. Machine learning and artificial intelligence provide new possibilities to automate PiMIX workflows and are therefore transformative. PiMiX is an update of an earlier workflow~\cite{r9}.}
  \label{fig:2}
 \end{figure}

PiMiX consists of hardware, data and algorithmic components, which may be integrated using DDMs such as machine learning, as shown in FIG.~\ref{fig:2}. The PiMIX framework has three coupled loops towards DF: an experimental loop to collect data, a model loop to generate synthetic data for experimental data interpretation. The data fusion itself is usually an iterative process for experimental data and  modeled or synthetic data. Uncertainties and errors can come from all three loops. For the experimental loop, detector noise such as dark current and probabilistic interactions of photons and particles with objects and detectors give rise to signal fluctuations such as Poisson noise. For the model loop, in additional to computational errors, most kinetic or fluid models for a plasma can not make accurate predictions about the experimental data in ICF or MCF due to model uncertainties. One example of fluid model uncertainty is from the transport coefficients. For the data analysis, algorithmic uncertainties such as round-off errors are not avoidable~\cite{Neu:1961}. 

 Most ICF and MCF instruments generate raw data in 1D, such as neutron, charged particle or X-ray spectroscopy, or in 2D, such as X-ray, $\gamma$-ray, or neutron images. These raw data, after postprocessing, can be used to reconstruct 3D structures or time-dependent 3D structures of the plasmas, which contain rich physics such as implosion asymmetry in ICF, or structures due to energetic particles, plasma flows, waves, and turbulence in MCF. The 1D and 2D raw data generation can be described by the following universal equation,
 \begin{equation}
\mathscr{Y} = M \mathscr{X} + \mathscr{B},
\label{eq:em1}
\end{equation}
where $\mathscr{Y}$ is the 1D or 2D raw data from the measurement, $\mathscr{B}$ is the measurement noise, such as the dark field in imaging. $\mathscr{X}$ is the plasma property such as X-ray or neutron emissivity. $M$ is the measurement matrix. In most cases encountered in fusion experiments, $M$ is assumed to be a function of geometry and detector materials properties, which do not depend on plasma properties$\mathscr{X}$ and therefore Eq.~(\ref{eq:em1}) admits linear superposition of solutions to plasma emissivity. 
The linear superposition property also allows $\mathscr{X}$ to be expressed in terms of a superposition of a complete set of basis functions $\psi_n$,
 \begin{equation}
 \mathscr{X} = \sum_n c_n \psi_n.
 \end{equation}
Besides Fourier and wavelet decomposition, other recent examples for 2D basis function method for X-ray and neutron imaging applications are Laguerre-Gaussian modes decomposion in 2D, and spherical harmonic decomposition in 3D~\cite{Vol:2017}. 

Finding $\mathscr{X}$ from $\mathscr{Y}$ therefore reduces to the well-known linear inverse problem or linear matrix inversion when $\mathscr{X}$ and $\mathscr{Y}$ are discretized (digitized as in 1D spectra or 2D images or 3D voxel arrays). Traditional algorithms such as Gaussian elimination require that the number of equations matches the number of unknowns. The number of unknowns in $\mathscr{X}$ (a matrix after discretization) can be quite large, which correspondingly requires a large number of measurements $\mathscr{Y}$, a large known measurement matrix $M$ and well characterized noise properties $\mathscr{B}$. The number of unknowns in $\mathscr{X}$, $N(\mathscr{X})$, may be estimated to be $(L/\delta)^{D}$, with $L$ being the size of the plasma, $\delta$ the spatial resolution and D = 2 (two dimensional), or 3 (three dimensional). To resolve a 1-mm of an ICF implosion target, for example, a 2D model for $\mathscr{X}$ would require $N(\mathscr{X})$ = 10$^4$,  10$^6$ at $\delta =10$ and 1 micrometer resolution, respectively. $N(\mathscr{X})$ in 3D model would become 10$^6$ and 10$^{9}$ at 10 and 1 micrometer resolution. When a back-lighter as an external radiation source for ICF diagnostics rather than the plasma emissivity is used, Eq.~(\ref{eq:em1}) can still be used. The plasma-induced backlighter attenuation and scattering are now described by the unknown $M$ with $\mathscr{X}$ for the source is assumed to be known. 

\begin{figure}[htbp] 
  \centering
   \includegraphics[width=3.5in, angle=0]{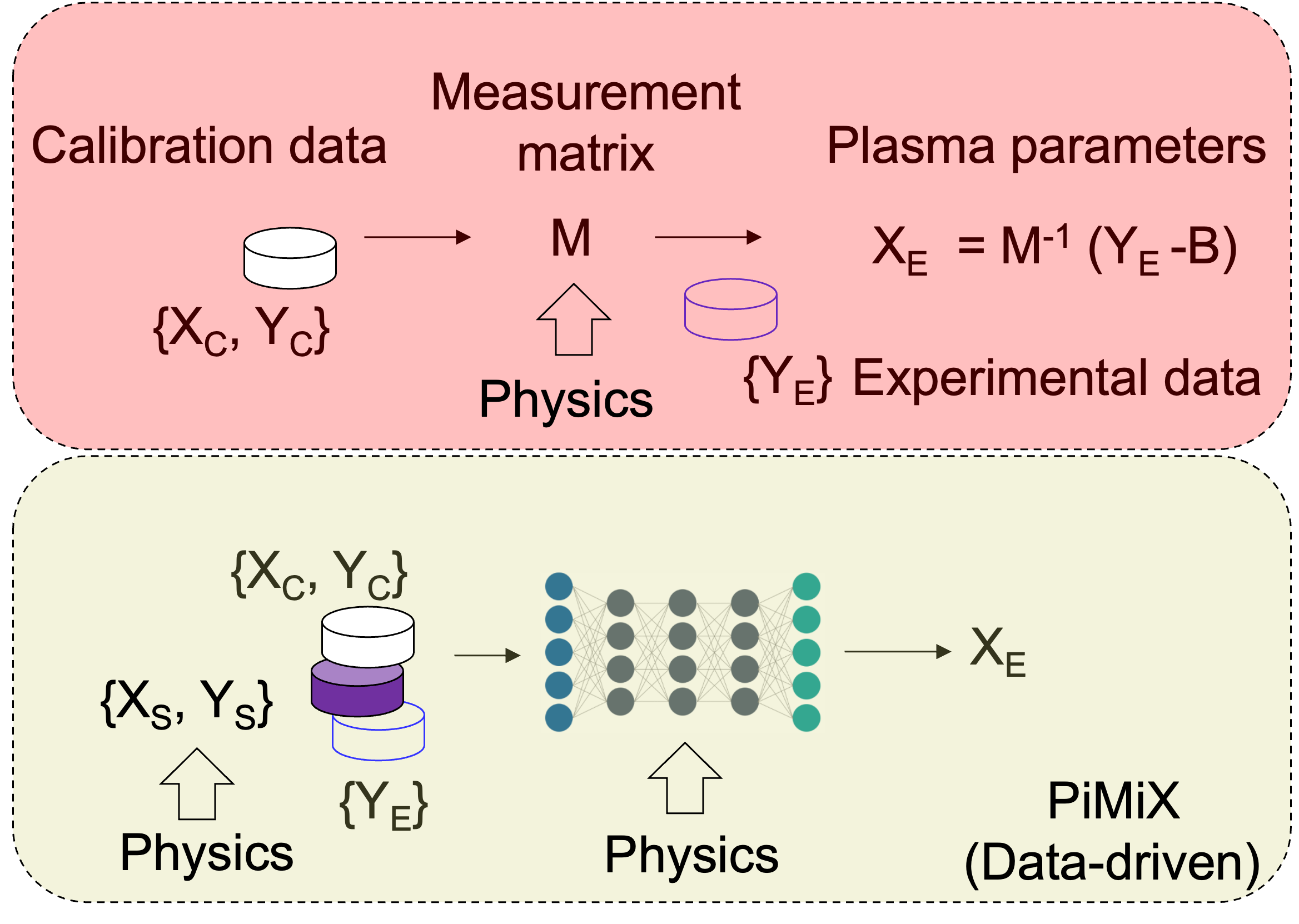} 
   \caption{A comparison of the traditional explicit-function-driven workflow (top) with the data-driven neural network (`implicit function') workflow in PiMiX (bottom). }
  \label{fig:pimix1}
 \end{figure}

There are a number of difficulties in implementing the traditional approach in practice. A large number of unknowns, depending on the desired plasma temporal and spatial resolution, require a correspondingly large number of calculations, which may be symbolically described by the inverse equation
\begin{equation}
\mathscr{X} = M^{-1} (\mathscr{Y} - \mathscr{B}) + \mathscr{X}_0,
\label{eq:em2}
\end{equation}
where $M^{-1}$ stands for the inverse matrix to $M$ if the inverse matrix exist, or pseudo-inverse if the inverse matrix is ill defined. Pseudo-inverse arises, for example, when the number of the known equations is less than the number of unknowns, or when $M$ is not a square matrix. One example of pseudo-inverse is Moore-Penrose pseudo-inverse. $ \mathscr{X}_0$ is an arbitrary vector in the NULL space of the measurement matrix, {\it i.e.} $M \mathscr{X}_0 = 0$.

A traditional PDM approach requires known measurement matrix $M$ and noise vector $\mathscr{B}$ through calibration, which we may call `explicit' calibration functions as in Fig.~\ref{fig:pimix1}. Detailed physics models and meticulous calibrations are usually required to derived measurement matrix $M$ and dark field vector $\mathscr{B}$. Furthermore, X-ray and neutron interactions with matter (including detector materials) are probabilistic and contain different interaction branches such as photo-electric (or neutron) absorption, coherent scattering and incoherent scattering. For high energy X-rays and $\gamma$-rays, the electron-positron pair production (or corresponding neutron-induced nuclear reactions) is also possible. In other words, for the same radiation dose, the measurement matrix $M$ can't strictly be considered to be a constant, even though this is often assumed. As was summarized in~\cite{r8}. Many factors can contribute to variations in $M$ for a specific measurement: spectral response, flux response, detection efficiency, temporal resolution, spatial resolution, and dynamic range. Similar considerations may be applicable to the dark field vector $\mathscr{B}$. In a fusion plasma environment, it can be difficult to know $\mathscr{B}$ ahead of an experiment. In short, high-resolution measurement of plasmas based on X-rays and neutrons requires correspondingly large number of measurements, calibrations, and background characterization. The probabilistic interactions of X-rays and neutrons with matter imply that the solutions to Eq.~(\ref{eq:em1}) and (\ref{eq:em2}) may not be unique for the same measurement, only a probabilistic interpretation of X-ray, $\gamma$-ray and neutron data, such as maximum likelihood interpretation, is possible. 

DDMs based on neural networks, especially deep neural networks (or neural networks with more than two layers of nodes), are now growingly adopted in 1D (spectroscopy) and 2D (imaging) data processing outside nuclear fusion and plasma physics. Here we discuss their applications in burning fusion plasma experiments, in particular, X-ray and neutron imaging through the PiMiX framework. Examples of deep neural networks include fully connected neural networks, convolution neural networks, autoencoders, encoder-decoder networks, generative adversarial networks, {\it etc}. Despite of the network architectural variations, workflows in using the neural networks for data processing are similar as shown in the bottom frame of Fig.~\ref{fig:pimix1}; {\it i.e.} each of the network requires a set of data, $\mathscr{Y}$ in Eq.~(\ref{eq:em1}) as inputs, and $\mathscr{X}$ as outputs. Solving Eq.~(\ref{eq:em1}) now turns into a.) identifying a neural network architecture; b.) selecting the activation function; and c.) tuning the parameters of the neural network including the weights of the neurons or nodes. One key difference between the DDMs, also called machine learning (ML) and artificially intelligent (AI) methods, and the traditional PDMs is that, the neural network architecture, instead of explicit functions $M$ and $\mathscr{B}$, is used in the data processing workflow. In other words, neural networks may be regarded as `implicit functions', which depend on the network architecture, neuron weights, and activation functions.  	

Although physics is not explicitly required in DDMs, which is sometimes perceived as a `blackbox magic' of machine learning, there is a growing number of ways that physics may be incorporated, including but not limited to: 1.) Selecting physics motivated variables and their pairing $\{\mathscr{X}$, $\mathscr{Y}\}$; 2.) Selecting datasets for inputs and define outputs; 3.) Selecting neural network architecture, activation function and other hyper-parameters; 4.) Using physics models to generate synthetic data sets, $\{\mathscr{X}_S, \mathscr{Y}_S\}$ as shown in Fig.~\ref{fig:pimix1}; 5.) Enforcing physics and constraints in neural network loss functions; and 6.) Assessing and validating the results generated by DDMs. Incorporating physics into development of DDMs may be called `physics-informed' or `physics-regulated' machine learning and AI. Adding physics may also decrease the amount of experimental data needed for deep learning.

\begin{figure}[htbp] 
  \centering
   \includegraphics[width=3.5in, angle=0]{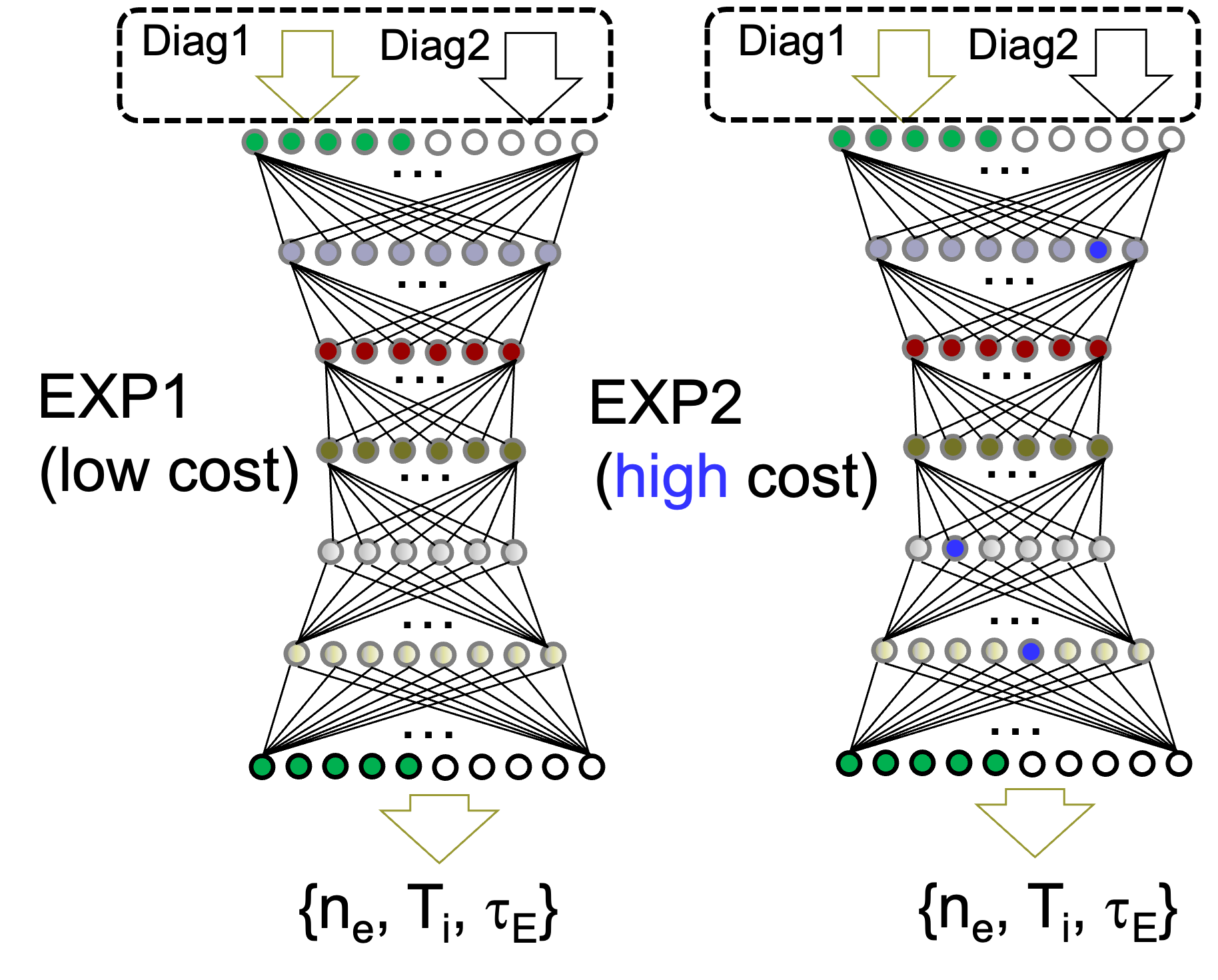} 
   \caption{Data augmentation is an important part of PiMiX strategy when the experimental data alone are not sufficient for neural network training. We may use low-cost experimental emulation and surrogates rather than synthetic data from simulations in PiMiX. Tuning and validations by new surrogate experiments can be readily designed and deployed at, for example, a light source facility like the Argonne Advanced Photon Source. Transfer learning is then used to deploy the PiMiX to nuclear fusion experiments.  }
  \label{fig:Trans1}
 \end{figure}

DDM approaches to PiMiX allow integration of multiple diagnostics or `meta-instruments' that merge hardware not physically collocated. Integrated measurement capability is a new frontier in fusion instrumentation, enabled by the growing number of instruments and data, and DDMs such as deep learning. As shown in Fig.~\ref{fig:Trans1}, data from two or more diagnostics may be fed into a neural network to generate physical quantities of interest. Here we assume that model- and physics-motivated quantities, such as plasma density ($n_e$), ion temperature ($T_i$), and energy confinement time ($\tau_E$), neutron yield ($Y_n$), are a function of the signals from Diag1 and Diag 2, $n_e = f (Diag1, Diag2)$. Explicit form of the function $f$ is not required. A neural network correlates the input data with $n_e$, $T_i$, and $\tau_E$ implicitly, based on physics considerations. Some additional discussions are given in Sec.~\ref{sec:phys}.

DDM approaches to PiMiX also allow data augmentation to compensate for insufficient datasets. It has been recognized that in many fusion experiments including ICF, there are insufficient amount of experimental data for training of deep neural networks from scratch. Transfer learning can be used for data augmentation~\cite{PY:2010}. A transfer learning workflow has been described for ICF experiments~\cite{Hum:2020}. The additional benefits include simulation calibration, low fidelity simulations validation against high fidelity simulations. In Fig.~\ref{fig:Trans1}, we propose to use `low cost' high data-throughput experiments to augment data from `high-cost' low data-throughput experiments. While X-ray instruments have demonstrated nanometer resolution under certain conditions through long signal integration time, repetitive measurements, and the use of monochromatic light, this feat may be hard to reproduce in transient ICF experiments, when the experiments last less than a fraction of one millisecond and are very costly to reproduce. An X-ray and neutron imaging spatial resolution in the range of 1-10 micrometers would be superior to the state-of-the-art in ICF~\cite{r8}. The sub-ns resolution is dictated by the need to produce movies of implosion dynamics that can be directly compared with the state-of-the-art hydro codes that run on today’s best high-performance computers. 
 
 Even though we emphasize the use of the ML and AI methods for PiMiX here, we may combine ML and AI methods with traditional PDMs together for better results. Such a combination may be necessary for physics-interpretable ML for fusion applications. For example, it is straight forward to expand Eq.~(\ref{eq:em1}) to include two measurement matrices, one each for Diag1 and Diag2 as shown in Fig.~\ref{fig:Trans1}, such and similar PDMs would complement and enhance DDMs, and {\it vice versa}, as discussed in Sec.~\ref{sec:stat}. 

 \subsection{Statistical inference and optimization in PiMiX \label{sec:stat}}
 
Due to uncertainties in models (including ML models), computational errors, instrument noise, random yet significant background in fusion plasma experiments, and probabilistic interaction physics of X-rays and neutrons, probabilistic inferences need to be applied to both traditional PDMs, such as Eq.~(\ref{eq:em1}), and data-driven models, such as deep neural networks or deep learning. 

The uncertainties, noise and errors turn the exact solutions of Eq.~(\ref{eq:em1}) into an statistical inference of optimized estimate for $\mathscr{X}$, 
\begin{equation}
\mathcal{L} (\mathscr{X} ) \equiv \| \mathscr{Y} - M \mathscr{X}  \|_p  < \epsilon.
\end{equation}
 Here we introduce the loss or cost function $\mathcal{L} (\mathscr{X})$ as the difference between  $\mathscr{Y}$  and  $M \mathscr{X}$. $\epsilon$ is the sum of all noise and errors, which may give the upper bound to  $\mathcal{L}$. The loss function is calculated by using the so-called $l^p$ norm~\cite{BV:2009}. If $p = 2$, the difference is calculated by using the $l^2$ norm or the Euclidean  distance. In some cases, $p = 0$ may be desired, which is, however, computationally hard and related to an open Millennium Prize problem (P {\it vs.} NP). Additional details and references were given in Ref.~\cite{r9}. In practice, $p=1$ is often used.
 
  Coexistence of DDMs, PDMs and others (such as surrogate experiments) for $\mathscr{X}$ offers another approach to statistical optimization (`minimizing $\epsilon$') in PiMiX, which may be expressed in terms of the following equation in general~\cite{r9},
 \begin{eqnarray}
(\mathscr{X}, \tilde \mathscr{X}) & \equiv & {\rm argmin }_{f(\mathscr{X}) =0; \tilde f(\tilde \mathscr{X}) =0} \mathcal{L} (\mathscr{X}, \tilde\mathscr{X} ), \label{eq:op1}\\
\mathcal{L} (\mathscr{X}, \tilde \mathscr{X}) & \equiv &  \| I_{exp} (\mathscr{X}) - \tilde I_{syn} (\tilde \mathscr{X})\|_p +\mathcal{R} (\mathscr{X}, \tilde \mathscr{X}). \label{eq:op2}
\end{eqnarray} 
Here we generalize the cost function $\mathcal{L}$  to be the difference between a function of the experimental data such as images $ I_{exp} (\mathscr{X})$ (each is a 2D intensity map) and the corresponding synthetic data $  \tilde I_{syn} (\tilde \mathscr{X})$. $\tilde \mathscr{X}$ is the corresponding theoretical vector of $\mathscr{X}$. The regularization functions $f (\mathscr{X}) = 0$, and $\tilde f ( \tilde \mathscr{X})$ =0  could include energy conservation, rotational symmetry of the images, and others. Additional regularization to $\mathcal{L}$ through $\mathcal{R}$, such as the positiveness of the spectroscopy and imaging signals~\cite{Wang:2000}, is often used as well. 

Bayesian deep learning, Gaussian process and others have emerged as probabilistic deep learning frameworks to integrate deep learning with probabilistic models~\cite{WangH:2020}. Applications of probabilistic deep learning include statistical optimization of the inference results such as feature identification~\cite{Chen:2018}, uncertainty quantification, model and hypothesis testing~\cite{Sugi:2016}. Bayes by Backprop is an efficient and backpropagation-compatible algorithm for learning a probability distribution on the weights of a neural network~\cite{Blundel:2015}. Adoption of these existing probabilistic deep learning algorithms for PiMiX, including fully connected neural networks, convolutional neural networks, autoencoders, and generative adversarial networks, is on-going.

\subsection{Neutron, X-ray and $\gamma$-ray emission models and measurements \label{sec:phys}}
\begin{figure}[htbp] 
  \centering
   \includegraphics[width=6.0in, angle=0]{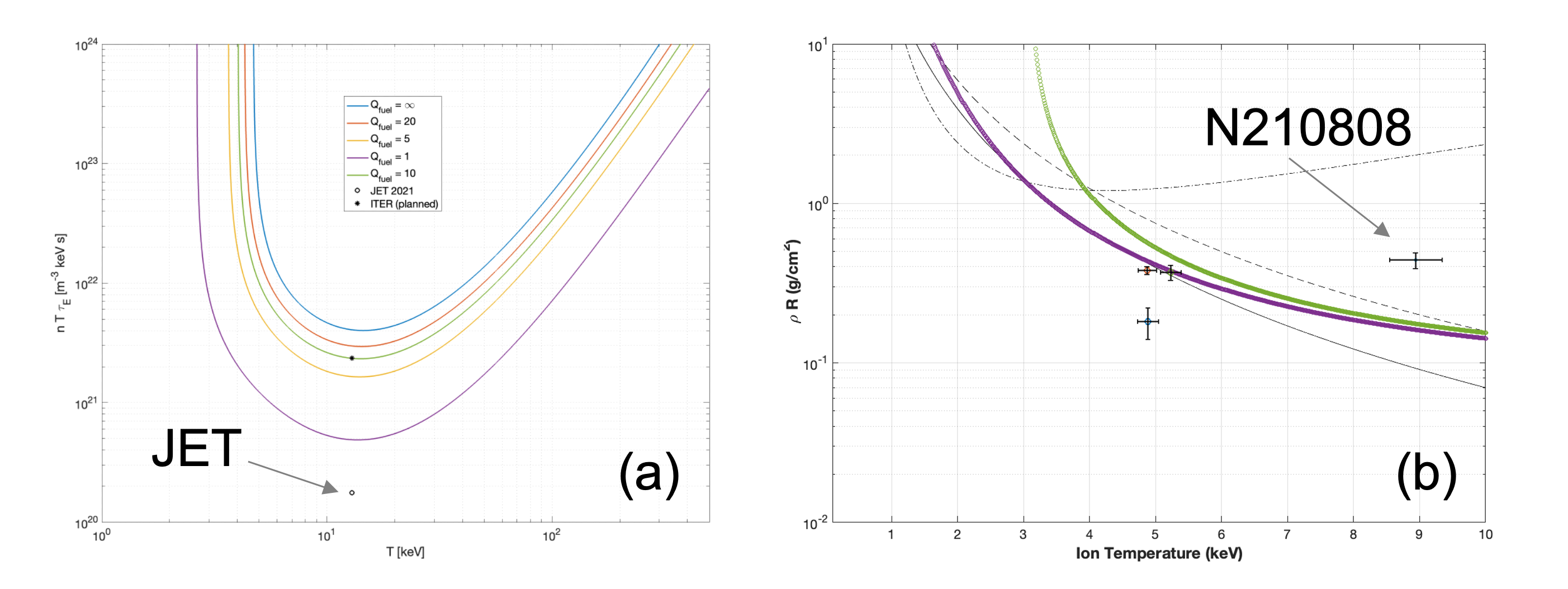} 
   \caption{Ignition boundary models in magnetic confinement fusion (a) and inertial confinement fusion (b), and comparisons with experiments in JET and NIF. For magnetic confinement, kinetic effects are ignored. For NIF, the residue kinetic energy (i.e., random motion on the hot spot during burn) is ignored. The imploding kinetic energy directly provides the heating energy of the DT fuel. The minimum required peak-no-burn temperature of hot DT (the T in Eq.~(\ref{Eq:Cheng1})) or hot DT pressure  strongly depends on the peak implosion velocity (see Eq.~(8) in Ref.~\cite{Cheng:2021}). The detailed derivation is given in~\cite{Cheng:2013}.}
  \label{fig:nif1}
 \end{figure}

Next, we discuss, as an example, the applications of PiMiX to assessment of ignition condition in thermonuclear fusion. The ignition condition for ICF and MCF is usually determined by energy and power balance, or the Lawson criterion~\cite{Law:1957}. In MCF, the Lawson criterion can be expressed in terms of the fusion triple product, $n T \tau_E$, as a function of temperature, see Fig.~\ref{fig:nif1}a, based on a recent tutorial paper~\cite{WS:2022}. 

Fundamental factors that affect ignition in ICF have been analyzed in details recently~\cite{Cheng:2021}. Additional models in light of the recent NIF ignition can be found in~\cite{Abu:2022, Cheng:2013,Cheng:2014,Bak:2022}. The condition for sustained thermonuclear burn of the hot spot in NIF is
 \begin{equation}
 \tau_{rep} \le \tau_H,
 \end{equation}
 where $\tau_{rep} = E_T/ P_{h}$ is the nuclear fusion reproduction time, and $\tau_H = R_{hs}/C_s^*$ is the hydrodynamic disassembly time. Plugging in the models for $\tau_{rep}$ and $\tau_H$ and assuming thermonuclear equilibrium for electrons and ions at different temperatures, one obtains~\cite{Cheng:2021}
 
 \begin{equation}
 \rho R \ge \frac{f_{DT} A_{DT}}{N_A} \frac{(3kT_i/2 + 3 kT_e/2 +E_{rad} /n_{DT}) C_s^* }{ \langle\sigma v \rangle W_f - (\dot{Q}^b +\dot{Q}^e) f_{DT}/n_{DT}^2},
 \label{Eq:Cheng1}
 \end{equation}
 where $n_{DT} = n_D +n_T$, $\displaystyle{f_{DT} = \frac{(n_D + n_T)^2}{n_D n_T}} $. Additional detailed definitions of the parameters in Eq.~(\ref{Eq:Cheng1}) are given in Refs.~\cite{Cheng:2013, Cheng:2014, Cheng:2021}. The different model predictions are shown in Fig.~\ref{fig:nif1}b. While some models can be readily eliminated based on the recent NIF ignition experiments, such as the shot N210808, other models may require most sophisticated measurements for higher resolutions and advanced data-processing framework such as PiMiX.

The suggested observables include~\cite{Cheng:2021}:   (1) a
nuclear energy reproduction time shorter than (or at least
close to) the hydrodynamic disassembly time; (2) a sharp and
narrow peak neutron flux so that the full burn width of the
capsule is close to the full-width half maximum (FWHM) as demonstrated in Figure 3 in Ref.~\cite{Cheng:2021}
(right); (3) the FWHM would be less than 50 - 60 ps for the current NIF capsule designs; and (4) a burn fraction close
to 1.65\%. A burn fraction of 3.3\% in a capsule with 180 microgram fuel
would produce 7.1 $\times$ 10$^{17}$ neutrons.

We summarize measurement capabilities and gaps to meet the physics model validation and fusion yield optimization requirements in Table~\ref{Tb:Signal}.  Quantitative neutron and high-energy photons (nuclear fusion $\gamma$-rays) measurements are especially important to further optimization of experiments such as NIF.
 
\begin{table}[htp]
\caption{Multi-modal measurement (1D and 2D) options for PiMiX, and a comparison of detectors, spatial resolution ($\delta$), data types, model for data processing, and 3D reconstruction algorithmic options.}
\begin{center}
\begin{tabular}{llccccc}
\hline
{\bf Modality}  & \multicolumn{4}{c}{Uni-modal} & Multi-modal\\\hline
&X-rays  & neutrons & $\gamma$-rays & charged & (PiMiX) \\ 
&&&&particles& \\
  & &  &  & [p/D, e$^-$/e$^+$] & \\ 
\hline\hline
 $\mathscr{X}$ & $\rho$R, $T_e$ & $T_i$, Y$_n$ & $\rho$R & $\rho$R & multi-physics  \\
&&flow, B field & $Y_n$ &B field& \\
$\mathscr{Y}$ & $I_S$, $E_X$ & $I_S$, nToF & $I_S$ & $I_S$, $E_C$ &  (multiple) \\
 &&&&& \\
Detector &  \multicolumn{5}{c}{$\longleftarrow \cdots $ scintillator + imaging sensor $ \cdots \longrightarrow$ } \\
 & & & Cherenkov & & (ultrafast Si)  \\
$\delta$ &  \multicolumn{4}{c}{$>$ 10 \textmu m} & (1-10 \textmu m)  \\
Data & \multicolumn{4}{c}{homogeneous} & heter. \\
Model& \multicolumn{4}{c}{explicit} & implicit \\
3D & Radon& Multi-view && Abel & Sparse \\\hline
\end{tabular}
\end{center}

\label{Tb:Signal}
\end{table}%

\section{PiMiX hardware \label{sec:hd} }
Extending upon the summary Table~\ref{Tb:Signal}, we first discuss detector hardware with an emphasis on ultrafast complementary metal-oxide-semiconductor (CMOS) image sensors, scintillator convertors, and imaging modalities with an emphasis on X-ray photon and neutron counting methods~\cite{r9,Wang:2015}.

\subsection{Ultrafast CMOS image sensors}
Recent advances in imaging sensors offer a growing number of choices for implementation in PiMiX, see Table~\ref{Tb:CMOS} for some examples. Many of the existing imager options are available through advances outside nuclear fusion. CMOS image sensors are rapidly replacing charge-coupled devices (CCDs) sensors as the primary digital image sensor technology for fusion energy and other applications~\cite{Lin:2023}. The state-of-the-art commercial fast CMOS cameras such as Phantom TMX 7510 have a full resolution of 1280 $\times$ 800 pixels at 76k frames per second (fps). In comparison, one of the highest frame rate burst-mode image sensors, reported by Tohoku University, Japan, has reached 100 Million fps with 368 frames record length. The key to this burst-mode high-speed performance was the high-density memory process, which formed deep-trench high-density capacitors inside each pixel to store the frames. For scientific instrumentation and niche market, such a unique fabrication process would be either not accessible or too expensive to implement. Fossum group in Dartmouth, in collaboration with LANL, is currently developing a high-speed continuous-wave (CW) sampling mode image sensor based on a low-cost 180-nm process (TMX 7510 used 110-nm process)~\cite{r6,Dartmouth}. With a superblocks-wise low-noise readout structure, a novel asynchronous SAR ADC design, and pump-gate global-shuttle pixel, the new Dartmouth sensor can run at 78.1~kfps but with much lower total noise than TMX 7510. The sensor will have 12 identical superblocks. Each superblock will be self-contained each with its own pixel array, readout circuits, and peripheral supporting circuits. Multiple superblocks can be instantiated on the same chip to configure the sensor resolution easily without sacrificing the frame rate. A 96$\times$120 test chip has been designed and taped out~\cite{Dartmouth}. Some of the open questions for high-speed CMOS applications in fusion include transient electromagnetic pulse (EMP) responses, radiation and neutron hardness.

\begin{table}[htp]
\caption{High-speed silicon (Si) CMOS image sensor options for PiMiX, additional details including references in~\cite{r9}. CW stands for continuous-wave sampling mode.}
\begin{center}
\begin{tabular}{llccccc}
\hline
Technology  & Sensor  & framerate & No. of frames & Pixel & Dynamic \\
 &thickness (\textmu m)& (MHz) & (burst mode) & pitch (\textmu m) & range \\ \hline\hline
UXI & $<$ 100 & 500 & 4 & 25 & $\sim$ 10$^3$  \\
pRad2 & $<$ 100 & 4 & 10 & 40 & $<$ 10$^3$  \\
AGIPD & 500 & 4.5+ & 352 & 200 & 10$^4$  \\
KeckPAD & 500& 6.5 & 8 & 150 & $>$ 10$^3$   \\
ePix100 & 500 & 120 Hz & CW & 25 & $>$ 10$^3$  \\
EIGER &450 & 9 KHz  & CW & 75 & $>$ 10$^3$  \\
Ultra UBSi & $<$ 100 & 10$^3$ & 25 & 7.4 & $<$ 10$^3$ \\
Teledyne (Dalsa) & $<$ 100 & 100 & 16 & $<$ 50 &  \\
Dartmouth & $<$ 100 & 20 & 24 & 25 & $>$ 10$^3$ \\\hline
\end{tabular}
\end{center}

\label{Tb:CMOS}
\end{table}%

\subsection{Ultrafast Scintillators}
Scintillators are widely used in nuclear fusion and plasma experiments for X-ray, neutron detection, and radiographic imaging using penetrating ionizing radiations such as relativistic protons and electrons. However, most instruments are limited by scintillators that were developed decades ago, which can't meet the increasingly demanding experimental requirements such as in ignited plasma environment. Two new classes of scintillators are emerging that could significantly outperform the traditional materials. One is ultrafast inorganic scintillators, which are also pursued in parallel by the high-energy physics (HEP) community for experiments such as the Large Hadron Collider (LHC) and high-luminosity LHC (HL-LHC). Table~\ref{Tb:scint} lists optical and scintillation properties of several promising inorganic scintillators, characterized at the Caltech HEP Crystal Lab~\cite{Cal1,Cal2}. Another class of scintillators is organic-inorganic lead halide perovskites with ABX3 structure. The new pervoskites have attracted attention in recent years for ionizing radiation detection due to their short attenuation lengths, high stopping powers, large mobility-lifetime products, tunable bandgaps, simple and low-cost single crystal growth from liquid solution processes. Furthermore, advances in nanofabrications have resulted in novel surface structures that overcome the refractive index mismatch and can direct light to image sensors for higher detection sensitivity than conventional scintillators without using such meta-structures.

\begin{table}[htp]
\caption{Fast and ultrafast inorganic scintillators and their properties characterized at Caltech HEP Crystal Laboratory~\cite{Cal1, Cal2}. Organic scintillators, which may be better for neutrons may be found in Ref.~\cite{Wang:scint}. Light yield is within the 1st ns.}
\begin{center}
\begin{tabular}{llccccc}
\hline
Scintillator  & BaF$_2$  &BaF$_2$:Y & ZnO:Ga & LYSO:Ce & GAGG:Ce \\
density (g/cm$^3$) & 4.89 & 4.89 & 5.67 & 7.4 & 6.5  \\
Melting point ($^o$C) & 1280 & 1280 & 1975 & 2050 & 1850  \\
X$_0$ (cm) & 2.03 & 2.03 & 2.51 & 1.14 & 1.63  \\
R$_M$ (cm) & 3.1 & 3.1 & 2.28 & 2.07 & 2.2  \\
Z$_{eff.}$ & 51.6 & 51.6 & 27.7 & 64.8 & 51.8  \\
dE/dX (MeV/cm)& 6.52 & 6.52  & 8.42 & 9.55 & 8.96 \\
$\lambda_{\rm peak}$ (nm) & 220 & 220 & 380 & 420 & 540 \\
Refr. Index & 1.5 & 1.5 & 2.1 & 1.82 & 1.92 \\
Light yield (ph/MeV) & 1200 & 1200 & 610 & 740 &  640\\
Decay time (ns) & 0.6 & 0.6 & $<$1 & 40 & 53  \\ \hline
\end{tabular}
\end{center}

\label{Tb:scint}
\end{table}%

\subsection{Imaging modalities}
\begin{figure}[htbp] 
  \centering
   \includegraphics[width=4.5in, angle=0]{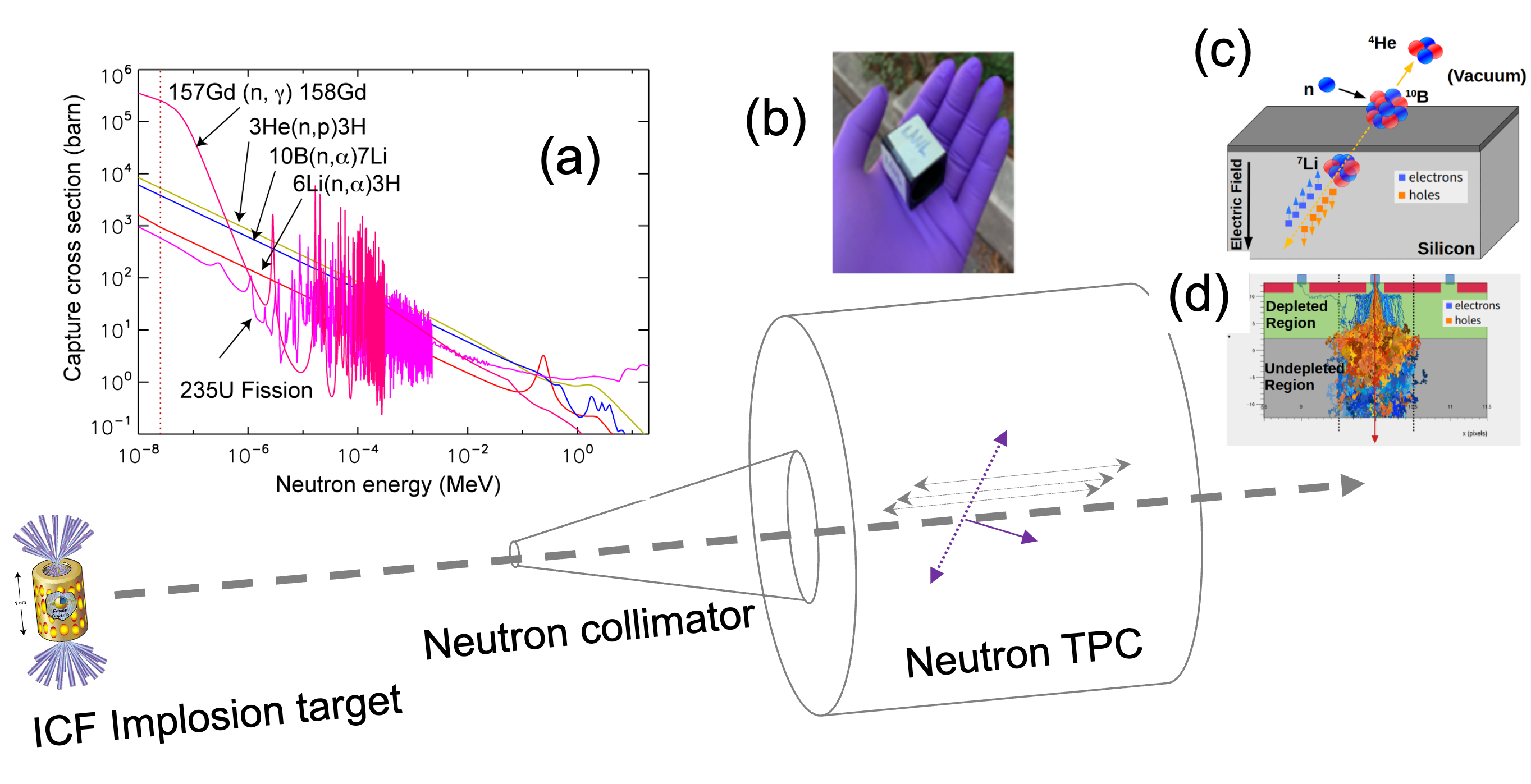} 
   \caption{A neutron TPC design for nuclear fusion applications. The incoming neutron momentum vector {\bf P}$_0$ can be made accurately, which allows for accurate neutron energy measurement through position-sensitive neutron or recoil particle detection as well as time-of-flight (ToF).}
  \label{fig:tpc1}
 \end{figure}

Based on the above discussions, time-resolved quantitative measurements of X-ray, neutron and fusion $\gamma$-ray signals are highly desirable in characterizing the implosion process and further experimental optimization for higher neutron yield and complete burn of the DT fuel. While most of the existing measurements are based on neutron and X-ray flux integration, which tends to smear out energy, position and other important information from individual energetic photons and neutrons.  We recently described a neutron time-projection chamber (TPC) design, which can significantly mitigate the information loss in the measurement of neutrons~\cite{Wang:2022O}. Fig.~\ref{fig:tpc1} summarizes the conceptual design and the existing building blocks for neutron TPC such as the CMOS detectors, absorption and scattering elements for fusion neutrons, and the existing simulation framework based on Allpix Squared~\cite{r6}. Work is also ongoing to extend the use of the TPC to measure nuclear fusion $\gamma$-rays.

\section{Data and machine learning in PiMiX \label{sec:dataM}}
In this section, we summarize the progress in data sets collection, synthetic data generation aiming at better understanding/emulating experimental noise, improving the signal-to-noise ratio to assist feature identification and image labelling, and machine learning algorithms development for PiMiX. 

\subsection{Experimental data from different facilities}
\begin{figure}[htbp] 
  \centering
   \includegraphics[width=5.0in, angle=0]{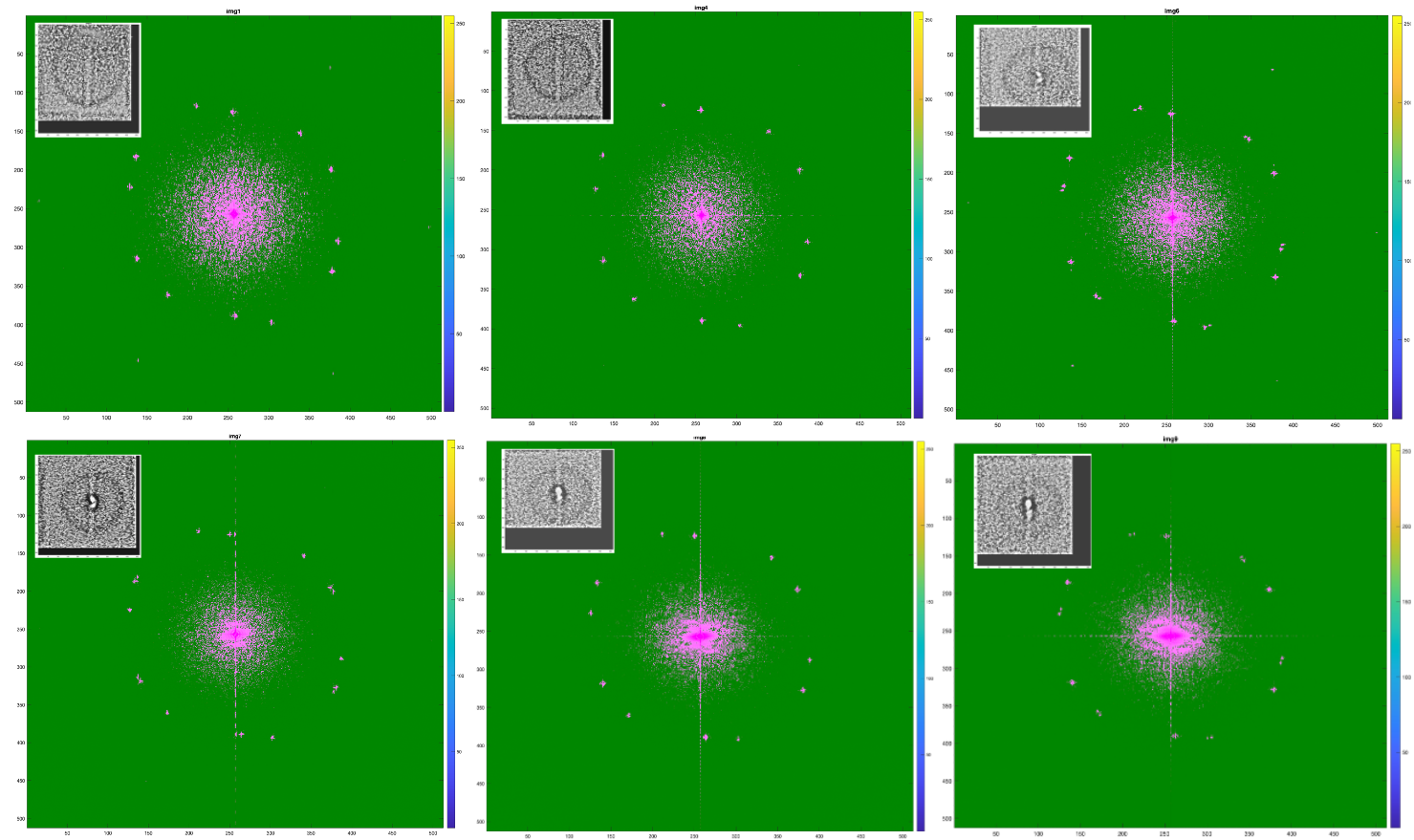} 
   \caption{An example of gated X-ray detector (GXD) implosion movie from NIF (Shot N180918). The black \& white (b\&w) inserts are X-ray intensity images. The six colored panels are Fourier transforms of the corresponding b\&w intensity images.}
  \label{fig:gxd1}
 \end{figure}

 NIF provides X-ray, neutron and fusion $\gamma$-ray experimental data that are generated at high cost and a low repetition rate. Such data are not reproducible else where, which limit in part the available experimental data for machine learning models.  An existing X-ray movie from NIF is shown in Fig.~\ref{fig:gxd1}. 

The total number of experimental NIF X-ray images exceeds 100 but is less than $\sim$1000. Through segmentation, 
transformations (rotation, for example) and other data augmentation techniques, it is possible to expand the experimental image set by a factor of $\sim$10. Some issues with the dataset are the low signal-to-noise ratio and low resolution for definitive feature recognition. The second issue with NIF X-ray images is that most of the images are not labelled, which limit their use for supervised machine learning.

\begin{figure}[htbp] 
  \centering
   \includegraphics[width=4.0in, angle=0]{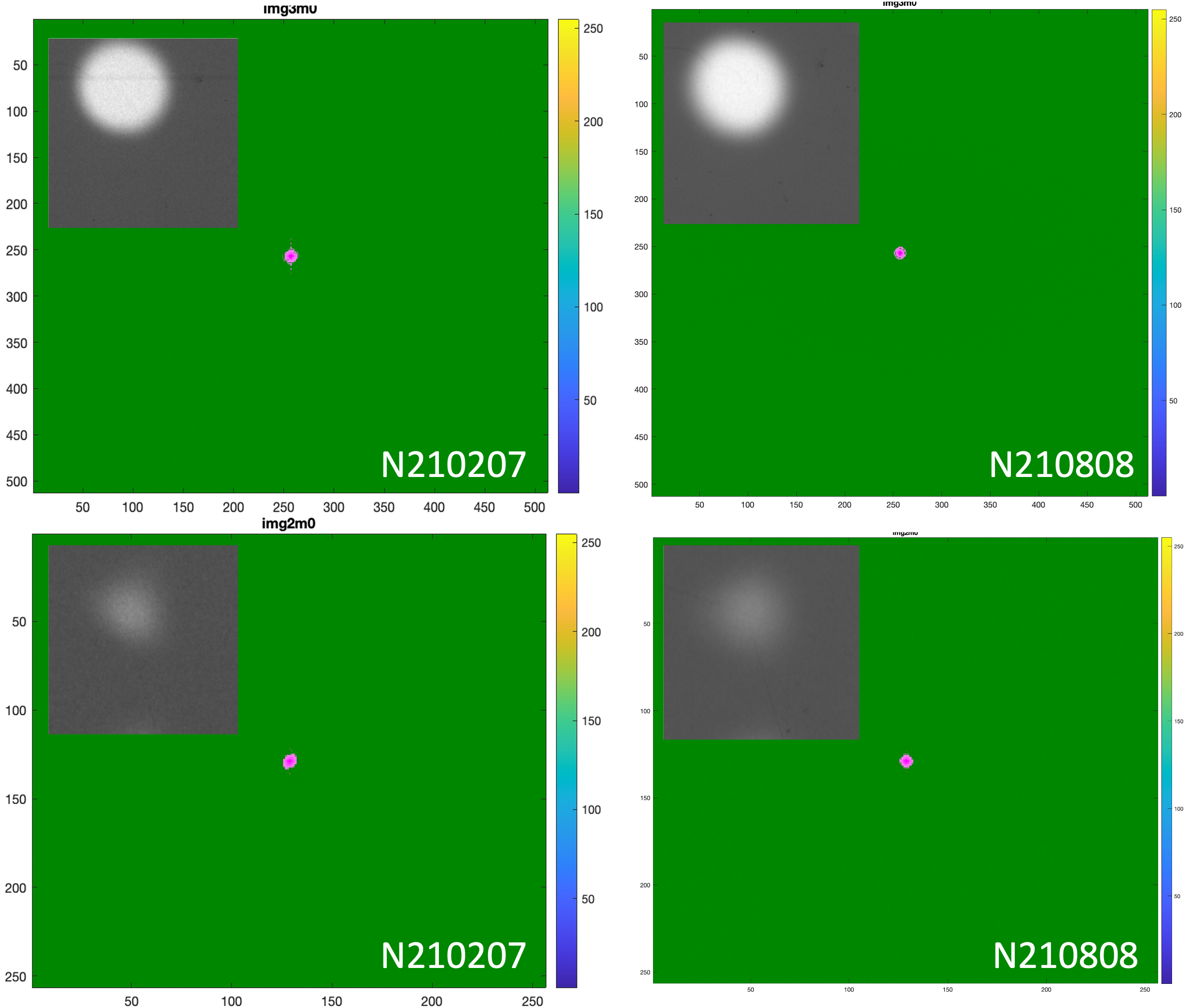} 
   \caption{A comparison two  sets of neutron images from NIF. The viewing location is at the equator 090-213. The b \& w inserts are neutron intensity images. The colored panels are Fourier transforms of the corresponding b\&w neutron intensity images.}
  \label{fig:nis1}
 \end{figure}

Two example sets of neutron data are shown in Fig.~\ref{fig:nis1} for shot N210207 (not ignited) and N210808 (ignited). The top row is for a pair of neutron penumbra images (non-ignition {\it vs} ignition), the bottom row is for the corresponding neutron images through a pinhole. The number of labelled neutron images is also very limited for ML purpose. 

 For experimental transfer learning, see for example Fig.~\ref{fig:Trans1}, and for multi-experiment DF (MXDF), PiMiX can leverage existing X-ray, neutron, and charged particle accelerator facilities for experimental data augmentation and algorithm development. User facilities such as OMEGA, OMEGA-EP at Rochester, and University of Michigan can generate X-ray, high-energy X-rays ($\gamma$-rays), charged particles, and neutron data by using high-power short pulsed lasers. The advanced photon source (APS) at Argonne and its upgrade provides/will provide synchrotron X-ray image data at 6.5 MHz (13 MHz for the upgrade). New detector capabilities through fast scintillators, metamaterial structures, and CMOS imaging sensors will be used to collect better experimental data for labelling and feature recognition. Existing NIF and new experimental data will be combined for transfer learning algorithm development and validation. In addition to facilitating reconstruction of 3D dynamic plasma movies, our goal is to achieve high-resolution density measurement in the range of one to ten micrometers, with sub-ns temporal resolution to produce the movies of implosion dynamics in NIF.

\subsection{Noise emulation \& statistical inferences \label{sec:NBIE}}

Improving signal-to-noise ratio for X-rays and neutron measurements, as shown in Fig.~\ref{fig:gxd1} and Fig.~\ref{fig:nis1}, is important to NIF and other fusion plasmas. Besides hardware methods, here we examine DDMs that are under development for PiMiX. In our earlier work~\cite{Wolfe:2021}, synthetic image generation did not sufficiently emulate the experimental noise. The noise was shown to significantly affect the downstream image-processing workflows such as 3D reconstructions from sparse X-ray data. NIF targets previously were approximated as constant density layers with perturbed interfaces. These perturbed interfaces were parameterized as Legendre polynomials in a half plane and the surfaces are produced by revolution. Coefficients of the polynomials were sampled from a distribution produced from measured values and individual samples were rejected based upon self-intersection of an interface and intersection between interfaces. Three dimensional volumes of the target were produced at a resolution of 512$^3$ voxels. NIST X-ray attenuation tables were used to produce a volumetric representation of the linear attenuation of the material at each voxel. To produce a synthetic radiograph, the python library TIGRE was used to ray-trace the voxel model given a source-object-detector geometry.

\begin{figure}[htbp] 
  \centering
   \includegraphics[width=4.0in, angle=0]{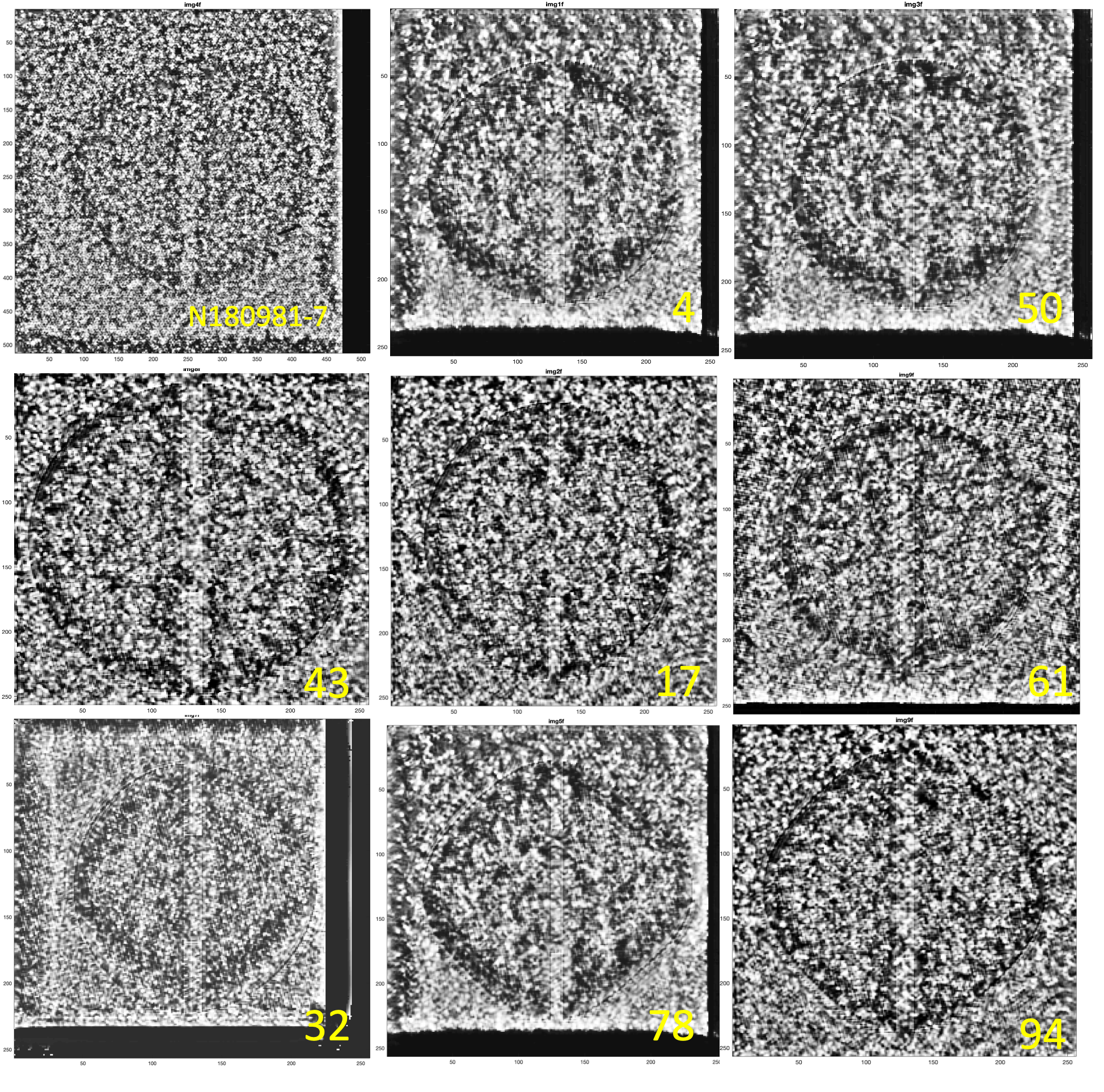} 
   \caption{Examples of X-ray synthetic data generated using a generative adversarial network in comparison with the NIF experimental in the upper left corner.}
  \label{fig:Wf1}
 \end{figure}

Generative ML and AI algorithms such as generative adversarial networks (GANs) are powerful tools for generating synthetic images that emulate experimental datasets, including noise, through unsupervised machine learning. Some examples of synthetic images that emulate the experimental X-ray images in NIF using Contrastive Unpaired Translation (CUT)~\cite{Park:2020}, an image-to-image translation tool, are shown in FIG.~\ref{fig:Wf1}. The key features and noise patterns of the synthetic images are qualitatively similar to the experimental image.

Experimental data are intrinsically imbedded with physics and materials information of particle interaction with a target. However, emulating experimental datasets using GANs are usually not transparent in terms of physics and imbedded materials information for synthetic images. To couple physics and material information more directly to neural networks, we also use existing multi-physics codes such as GEANT4, MCNP, and Bayesian inference engine (BIE) to generate synthetic data to augment experimental datasets. The Allpix Squared simulation framework~\cite{Spannagel:2018}, an open-source simulation tool that implements end-to-end simulations of particle detection from incident radiation to digitized detector output, was demonstrated to generate these datasets in neutron counting regime~\cite{r6}.

LANL developed a Bayesian Inference Engine (BIE) toolbox for synthetic radiographic image generation and experimental image analysis~\cite{Cun:1995}. Bayesian analysis of fusion plasma diagnostics has been reported before~\cite{Fischer:2003}. Machine learning has been recognized to potentially accelerate Bayesian methods for integrated diagnostic analysis (IDA) and applications such as real-time controls. BIE uses the object-oriented language Smalltalk for Bayesian inference, serving as a flexible tool for estimating unknown parameters in mathematical models by integrating prior information with observed data. In the context of radiographic measurement systems and geometric models for experimental objects, BIE enables the investigation of confidence intervals regarding the estimated object geometry and the comparison of likelihoods among different competing hypotheses.

\begin{figure}[htbp]
\centering
\includegraphics[width=0.45\textwidth]{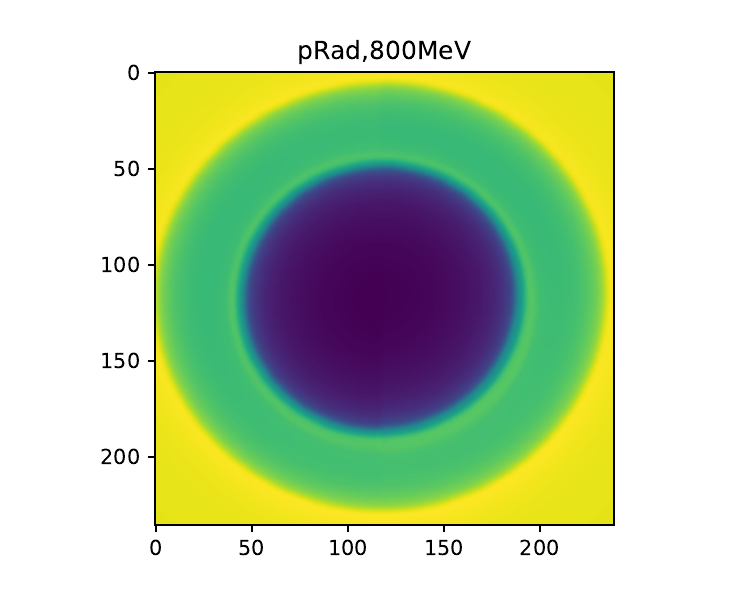}
\includegraphics[width=0.45\textwidth]{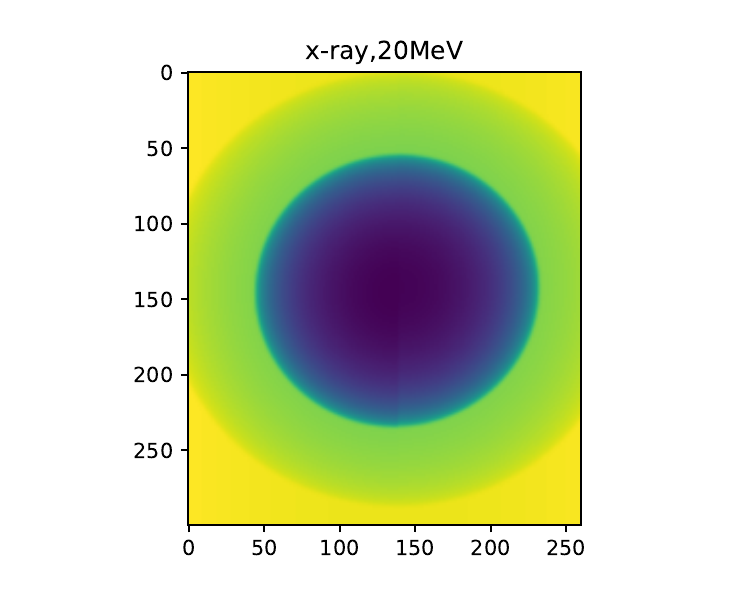}
\caption{The synthetic images using BIE depict the IQI using 800 MeV protons (left), and 20 MeV X-rays (right).} 
\label{fig:IQIsyn}
\end{figure}

 BIE consists of three components: a graphical programmer for the creation and manipulation of the measurement system model, a geometric modeler for the definition and interaction with the measurement system model, and an interactive optimizer. The engine not only derives meaningful estimates of model parameters but also offers a high degree of user engagement. By providing intuitive visualizations of the inference process and model behaviors, the BIE empowers users to grasp the intricate relationships between inputs, outputs, and parameter uncertainties. BIE is well-suited for scenarios involving nonlinear relationships between model parameters and image data, as illustrated in Figure~\ref{fig:IQIsyn} for a pair of X-ray and proton images of an Image Quality Indicator (IQI) object~\cite{Sjue:2020}. The IQI used here comprises a gold sphere with a radius ($r_2$) of 5 mm, enclosed by a
spherical shell made of Ti-6Al-4V alloy, extending to an outer radius ($r_1$) of 7.9 mm. To counteract energy
loss across the field of view and minimize chromatic aberrations in the proton measurement, we integrated a
dechromator which is made of polycarbonate cubic slab with sides of length 11.7 cm, that features a tapered
cylindrical hole at the center.
 
\subsection{Neural networks}

In PiMiX, different neural network and deep learning algorithms perform versatile tasks for DF workflows, complementing existing integrated data analysis (IDA) workflows using explicit models such as measurement matrices $M$. Prior to the recent demonstration of fusion energy gain $Q>1$~\cite{Abu:2022}, IDA has been used to validate physics models, to assess the efficiency of heating or implosion,  to mitigate instabilities, other energy and particle loss mechanisms, and to interpret complex data in fusion plasmas. In the ignited plasma era, transitions are ongoing from physics understanding, plasma characterization, model validation to fusion energy production and optimization ($Q> 10$ or more), from prototype materials to fusion power plant engineering, and from the first-wall physics to machine and environment safety. Adoption of both traditional PDMs and DDMs as in PiMiX to the new tasks such as integrated diagnostic-to-control
(IDC) is now timely and necessary. The growingly complexity, inter-disciplinary and multi-physics nature, and the harsh environment of fusion power plant operation also motivate PiMiX or traditional IDA for accelerated workflows, reduced measurement hardware through IDC, and real-time data handling capabilities for IDC.

We recently demonstrated deep neural networks for different plasma image processing tasks, as an essential ingredient of DDMs for PiMiX. A Kohonen neural network and a U-Net were used for tracking a large number of microparticles immersed in an ambient plasma~\cite{r3}. A deep convolutional neural network (CNN) was used for supervised classification of plasma images associated with kink instabilities~\cite{r4}. 3D reconstruction of ICF implosion image was possible by using a CNN employing an encoder-decoder architecture and sparse data~\cite{Wolfe:2021}. The 3D reconstructed features were found to be sensitive to noise. The 3D reconstruction has since been extended to generative adversarial networks as a part of model-dependent reconstruction study~\cite{r5}.  Further development of DDMs for PiMiX will emphasize the coupling of PDMs with DDMs, uncertainty quantifications~\cite{r9}, de-noising, and better DDM interpretability. 

\section{Experimental validations \label{ex:val}}
We have demonstrated several ways to extract more information from images by using DDMs and physics-informed DDMs. Here we highlight X-ray and neutron counting measurements using physics-informed ML called a fully connected neural network, followed by initial results in using BIE for multiple instrument (proton and X-ray) data fusion. As mentioned above, neutron and X-ray counting measurements yield more information than intensity-integrated measurements: including energy and position information of individual neutrons and X-rays.

\subsection{Super position resolution for neutron counting \label{sec:neut}}
Position-sensitive detection of neutrons has been used to extract more information such as fast neutron energy and momentum from neutron data~\cite{Wang:2022O}. For demonstration of sub-pixel or `super position' resolution, neutron counting measurements were carried out using customized CMOS image sensors with a neutron conversion layer of $^{10}$B~\cite{r6,Lin:2023b}. The secondary particles $\alpha$, $^7$Li, and $\gamma$-rays, generated from the neutron absorption by  $^{10}$B, reach the charge-depleted silicon layer within the detector to generate electron-hole (e-h) pairs as signals. The generated charge signals will travel the charged depleted silicon layer to the charge collection capacitors. The collected electric charge signals are then digitized to form images of neutron hits, see examples given previously in Ref.~\cite{r6}.

\begin{figure}[htbp] 
  \centering
   \includegraphics[width=4.0in, angle=0]{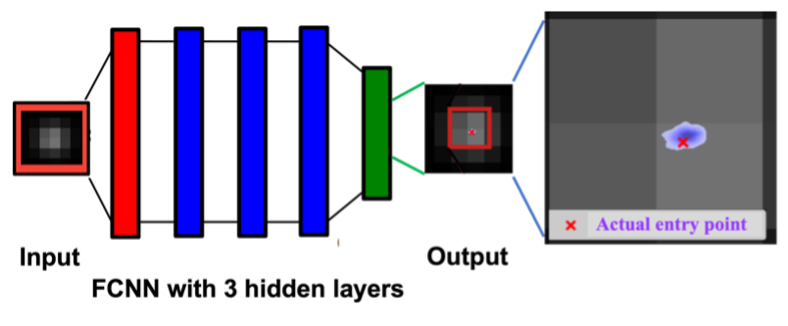} 
   \caption{A demonstration of super position resolution by using a physics-informed 5-layer fully connected neural network (FCNN) for neutron image processing. The FCNN were trained using physics-informed synthetic data generated from Allpix Squared~\cite{r6}.}
  \label{fig:Linn1}
 \end{figure}
 
 A five-layer fully connected neural network (FCNN) for neutron image processing is shown in Fig.~\ref{fig:Linn1}. This simple workflow was sufficient since the signal-to-noise ratios of the neutron-absorption signals were high and pre-processing of the images was not needed. Each neutron images had a cropped input size of 14$\times$14 pixels. Dropout layers, not shown explicitly in Fig.~\ref{fig:Linn1}, are also included in the FCNN during the network training and validation. Dropout layers helped to mitigate over-fitting issues as well as uncertainty quantification. The Allpix Squared code generated a synthetic dataset of 60k neutron events and their corresponding ground-truth labels for the neutron position. The physics regularization included the total energies of the secondary particles generated by neutron absorption. The blue kernel density estimation (KDE) plot in Fig.~\ref{fig:Linn1} confirms sub-pixel position resolution for neutron detection with the associated uncertainties.

\subsection{Energy-resolved X-ray measurements}

\begin{figure}[htbp] 
  \centering
   \includegraphics[width=4.0in, angle=0]{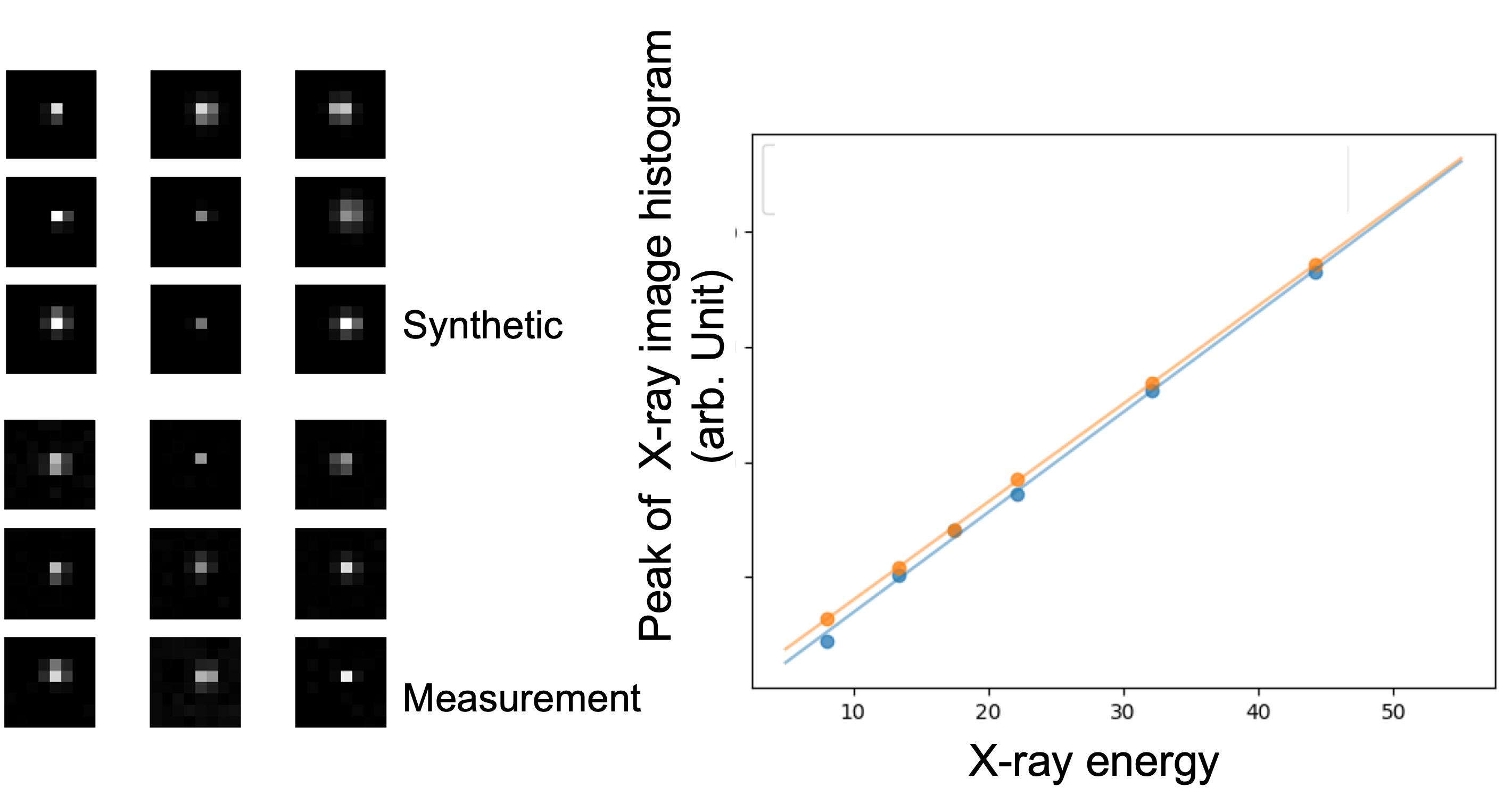} 
   \caption{X-ray energy calibration for synthetic data validated against experiments using an variable energy X-ray source. One the left, 9 synthetic X-ray images generated by Allpix Squared were compared with 9 examples of experimental images below. On the right, the energy information has been extracted from both the synthetic and experimental images, and were found to be consistent with each other.}
  \label{fig:fB1}
 \end{figure}

We extend the Allpix Squared data synthesis and analysis to X-rays in the photon-counting regime, Fig.~\ref{fig:fB1}. The experimental data were obtained using a variable energy X-ray source (8-60 keV), as described in~\cite{Wang:2016}. Various CMOS image sensors were tested for different X-ray detection efficiency and signal-to-noise optimization.

Even though the signal-to-noise ratio for X-rays were significantly lower than for neutron-induced signals as described in Sec.~\ref{sec:neut},  we still confirmed that X-ray energies can be resolved using the CMOS image sensors operated at room temperature. These initial results validate the use of Allpix Squared synthetic data for ML algorithm development and applications. The same detectors and data processing algorithms will be used for X-ray photon counting and time-integrated measurement with variable X-ray fluxes in plasmas and other related experimental settings.
 
 \subsection{Multi-instrument data fusion (MIDF)} 
\begin{figure}[htbp]
\centering
\includegraphics[width=0.4\textwidth]{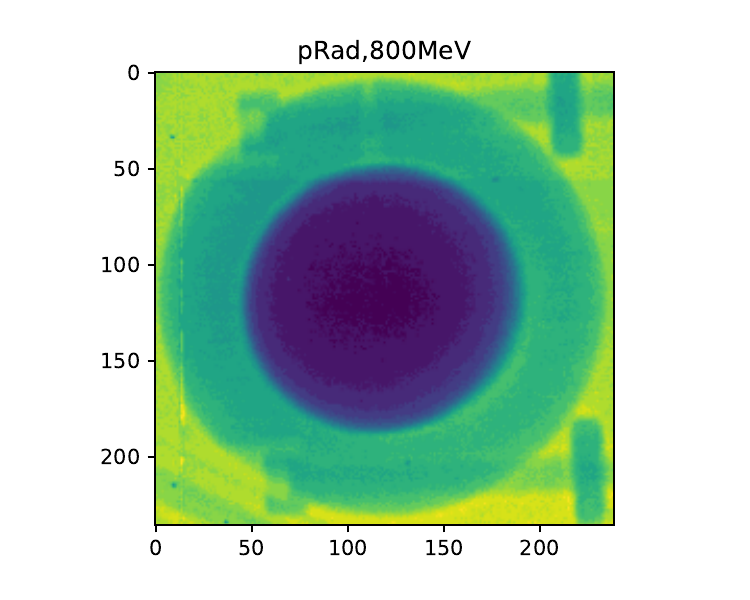}
\includegraphics[width=0.4\textwidth]{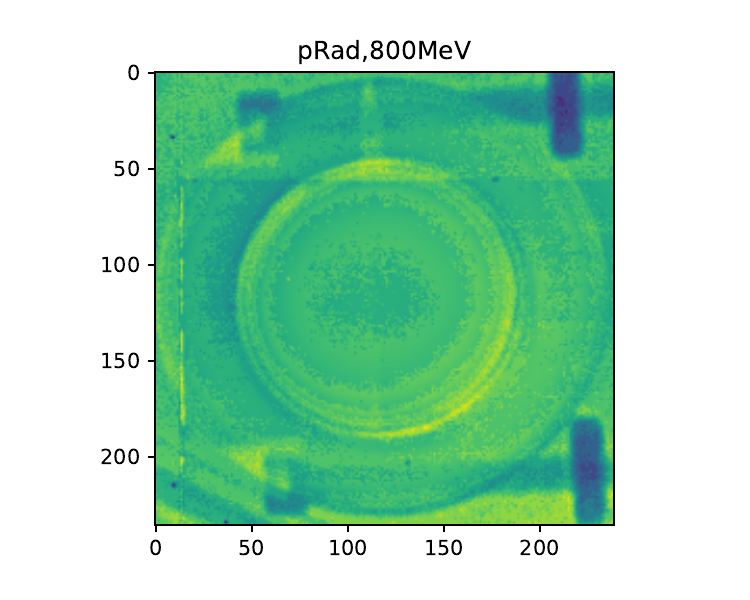}
\includegraphics[width=0.4\textwidth]{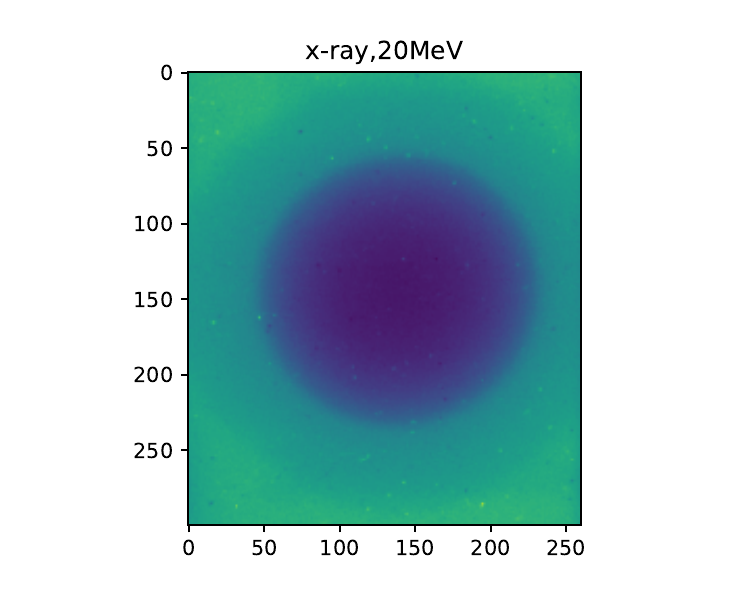}
\includegraphics[width=0.4\textwidth]{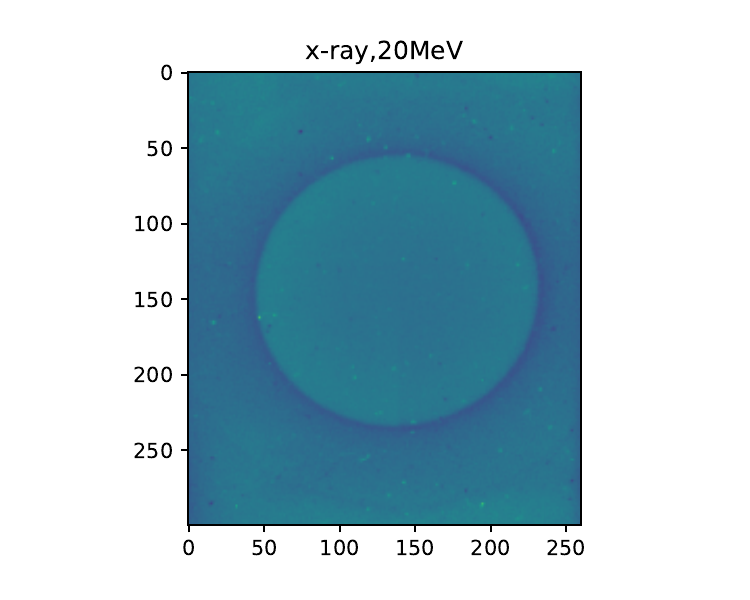}
\caption{The experiment images of the IQI using 800 MeV protons (top left), and 20 MeV X-rays (bottom left), and the corresponding chi-square differences, right column, between the experimental and synthesized images in Fig.~\ref{fig:IQIsyn}. The result indicates the feasibility of multi-instrument data fusion for better results than individual measurements alone.} 
\label{fig:IQIval}
\end{figure}

 To validate MIDF workflow and assess the quality of the synthetic images described in Sec.~\ref{sec:NBIE}, we applied the existing proton (800 MeV) and high-energy X-ray (20 MeV end-point energy) radiography imaging at LANL to collect experimental images from the IQI object. The experimental proton and X-ray image data, together with their comparisons with the synthetic images in Fig.~\ref{fig:IQIsyn} are shown in Fig.~\ref{fig:IQIval}.
 
Table~\ref{tab:ratio} summarizes the fitted radii and corresponding errors, calculated using the following equation,
\begin{equation}
    \delta(\frac{r_1}{r_2}) = \frac{(\frac{r_1}{r_2})_{\rm fitting}-(\frac{r_1}{r_2})_{\rm model}}{(\frac{r_1}{r_2})_{\rm model}}.
\end{equation}
This metric measures and compares the accuracy of the radii fitting. 

\begin{table}[htbp]
  \centering
    \caption{The table summarizes the fitting results for each scenario, encompassing individual species, dual species, and triple species. $r_1$ denotes the outer radius of the IQI, while $r_2$ represents the inner radius of the IQI core. $r_1/r_2$ is the ratio between the outer radius and the inner radius, serving as a metric for assessing fitting uncertainties. $\delta (r1/r2)$ indicates the uncertainty between the fitting result and the designed model.}

    \begin{tabular}{@{}c|cccc@{}}
      \hline
      &Model & X-ray 20 MeV & proton 800 MeV &  X-ray  \\
      &      &             &       &  + proton        \\
      \hline\hline 
      $r_1$ & 0.79 & 0.8285 & 0.8117 &  0.8120 \\
      $r_2$ & 0.5 & 0.5003 & 0.4969 & 0.5006 \\
      $r_1/r_2$ & 1.58 &  1.6559 & 1.6336  & 1.6219 \\
      $\delta(r_1/r_2)$ & & 0.0481 & 0.0339  & 0.0265 \\ 
\hline 
    \end{tabular}
  \label{tab:ratio}
\end{table}

This initial study using static IQI objects demonstrates that employing a combination of X-ray measurement at 20 MeV and proton measurement significantly reduces the uncertainty ($2.65\%$) compared to separate measurements alone ($3.39\%$ for protons and $4.81\%$ for X-rays at 20 MeV). The MIDF approach notably enhances measurement sensitivity, although its efficacy is contingent upon factors like beam profile, target density, and detector performance, which need to be considered together through the PiMiX framework. Extension of the static targets to transient targets in ICF implosion and others warrants further work.
 

\section{Summary \& conclusions}
Here we describe a new artificial intelligence (AI) and machine Learning (ML)-enhanced measurement and data fusion (DF)  framework, called Physics-informed Meta-instrument for eXperiments (PiMiX) for data collection, data handling and automated data interpretation, in the era of laboratory ignited plasmas. PiMiX, based on neural networks and deep learning algorithms, integrates physics understanding with heterogenous data streams from experiments, multiscale multi-physics simulations, and metadata such as material properties for data interpretation. PiMiX also offers the predictive potential for experiments and fusion plasma concepts. 

Data-driven methods (DDMs), especially machine learning and deep neural networks, complement traditional explicit function, such as measurement matrices $M$, approaches for data processing and interpretation. Some known issues with measurement matrices $M$ include nonlinearity, degeneracy and singularity, which might be addressed by constructing reduced-order models (ROMs) of the measurement or the underlying experiment. Explicit models complement DDMs through various physics-driven mechanisms such as synthetic data generation and cross-model validation.

DDMs allow DF from multiple diagnostics and construct `meta-instruments' in PiMiX.  Besides automated and integrated data processing from multiple diagnostics suite, PiMiX intends to address whether, through multiple instrument DF (MIDF), and/or multiple experiment DF (MXDF),  such as a combination of low-cost and high-cost (such as NIF or ITER) experimental data, and other forms of DF such as simulation and experiment DF (SXDF), more information can be extracted than using individual diagnostics or a single experimental device/facility alone. Initial results from PiMiX in super-resolution, energy-resolution and multi-instrument integration pave the way towards new fusion reactor concept designs and optimization of fusion experiments.

{\it Acknowledgements} We wish to thank Ann Satsangi and Joe Smidt (LANL ICF program managers) for support of the work. This work is also made possible in
part by US Department of Energy (DoE) under the Contract No. 89233218CNA000001.

\end{document}